\documentclass[aps,prl,twocolumn,showpacs,amssymb]{revtex4-1}
\usepackage{graphicx}

\usepackage[latin1]{inputenc}
\usepackage[rightcaption]{sidecap}
\usepackage{amssymb}
\usepackage{amsmath}
\usepackage{color}
\usepackage{subfigure}
\usepackage{floatflt}
\usepackage{units}

\begin{document}
\title{Ultrafast transport of laser-excited spin polarized carriers
in Au/Fe/MgO(001)}

\author{Alexey~Melnikov$^{1,2}$}\email[] {melnikov@fhi-berlin.mpg.de} \author{Ilya~Razdolski$^{3}$}\altaffiliation{Present address: Radboud University Nijmegen, The Netherlands}
\author{Tim~O.~Wehling$^{4}$}
\author{Evangelos~Th.~Papaioannou$^{2}$}\altaffiliation{Present address: Uppsala University, Sweden}
\author{Vladimir Roddatis$^{5}$}
\author{Paul~Fumagalli$^{2}$}
\author{Oleg~Aktsipetrov$^{3}$}
\author{Alexander~I.~Lichtenstein$^{4}$}
\author{Uwe~Bovensiepen$^{2,6}$}
\affiliation{$^{1}$Physical Chemistry Department, Fritz-Haber-Institute of the Max Planck Society, Faradayweg 4-6, 14195 Berlin, Germany\\$^{2}$Physics Department, Freie Universit\"{a}t Berlin, Arnimallee 14, 14195 Berlin, Germany\\$^{3}$Physics Department, Moscow State University, 119991 Moscow, Russia\\$^{4}$Theoretical Physics Institute, University of Hamburg, Jungiusstr.~9, 20355 Hamburg, Germany\\$^{5}$NRC "Kurchatov Institute", pl. Akademika Kurchatova 1, Moscow, 123182, Russia\\$^{6}$Faculty of Physics, University of Duisburg-Essen, Lotharstr.~1, 47048 Duisburg, Germany}

%
%
%
%
%
%
%

\date{\today}

\begin{abstract}
A time domain approach to probe hot carrier-induced spin dynamics is demonstrated. The experiments are performed in epitaxial Au/Fe/MgO(001), where spin-polarized hot carriers are excited in the Fe layer by 35~fs laser pulses. They propagate to the Au surface where the transient spin polarization is detected by magneto-induced second harmonic generation.
Different energies of majority and minority hot carriers excited in the exchange-split Fe band structure lead to their spin-dependent lifetimes in Au. Accordingly, two spin-polarized current contributions which propagate superdiffusively at different velocities result in a spin current pulse of about 100~fs duration.
\end{abstract}

\pacs{72.25.Ba, 72.25.Fe, 73.23.Ad, 78.48.J-}

\maketitle


The quest for fast, high-density, non-volatile magnetic memory is key motivation for research in the field of spin dynamics in low-dimensional ferromagnetic structures \cite{Hillebrands2006}. In recent years magnetization dynamics have seen a remarkable development from classical precession on the nanosecond scale to the ultrafast limit accessible by femtosecond laser pulses \cite{kirilyuk_RevModPhys10}. These findings still challenge the understanding of ultrafast magnetism and insight into the underlying elementary processes even in simple systems like Fe, Co, Ni is just at the beginning: (i) Koopmans and coworkers consider spin-lattice interaction as one origin of femtomagnetism and have developed an empirical model based on spin-orbit mediated electron spin-flip scattering \cite{koopmans_PRL05, koopmans_NatMat10}. (ii) Bigot et al. \cite{bigot_NatPhys09} and Zhang et al. \cite{zhang_NatPhys09} suggest that the light field is directly involved in magnetization dynamics. (iii) Battiato et al. propose superdiffusive spin transport being responsible for the ultrafast demagnetization \cite{battiato_PRL10}, which couples the fields of ultrafast spin dynamics and stationary spin transport \cite{jansen_JPhysD03}. It was shown \cite{Malinowski_NatPhys08} that in spin valve structures spin-polarized carrier transport speeds up and enhances the demagnetization compared to single ferromagnetic (FM) layers. Without doubt there is a need for additional investigations and development of a profound microscopic understanding. Novel experimental approaches, like the one presented here, bear the chance to overcome limitations of established schemes, which have difficulties to separate photon-, electron-, and phonon-mediated effects on the magnetization under non-equilibrium conditions established after femtosecond laser excitation.

Our experimental approach is based on the front-pump-back-probe scheme successfully used for the investigations of hot carriers (HC) dynamics in gold \cite{Brorson_PRL87,Juhasz_PRB93}: HC were excited by a short laser pulse within the optical skin depth $\lambda^S_{Au}\approx$15~nm at the front side of a thin gold film and the transient reflectivity $\Delta R/R$ monitoring HC traversing the film was measured at the back side through a transparent substrate. The delay in the onset of $\Delta R/R$ gives a direct measure of the propagation velocity of HC.

\begin{figure} \centering
 \includegraphics[width=1\columnwidth]{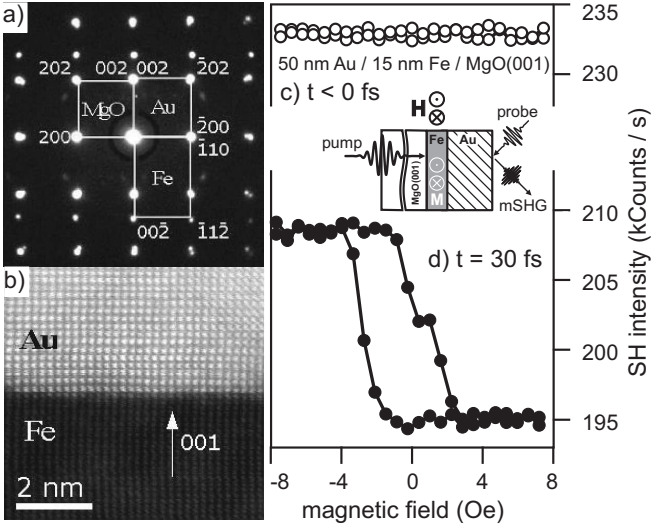}
\caption{\label{Fig1} Electron diffraction pattern of Au/Fe/MgO(001) (a), dark field STEM image of the Fe/Au interface (b), SH hysteresis loops at the Au surface before (c) and after (d) the excitation. The inset illustrates the pump-probe experiment.}
\end{figure}

To make the excited HC spin-polarized, a thin film of FM metal (Fe) of thickness $d_{Fe}\leq\lambda^S_{Fe}$ is introduced as shown in Fig.~\ref{Fig1}, inset. The \emph{pump} laser pulse focused onto the sample from the \emph{back} (substrate) side, excites HC predominantly in the exchange-split band structure of Fe due to $d_{Fe}\sim\lambda^S_{Fe}$. Owing to that, these HC may be \emph{spin-polarized} and, traversing the Au film, carry the spin polarization (SP) forming a pulse of \emph{spin current} (SC). Since there is no bias voltage applied, the charge component of this current might be screened by a displacement of "cold" carriers in Au due to the Coulomb interaction. This makes a dramatic difference to spin-polarized electric currents used for the spin transfer in spintronics \cite{Ralph_JMMM08} where the lateral size of devices is limited to $10-100~nm$ due to the charge current inducing the Oersted field proportional to the area of the device, which disturbs the experiment. The approach proposed here has the potential to overcome such limitations. The pulse of SC is detected at the Au surface by a magneto-optical signal from the \emph{probe} laser pulse focused onto the sample from the \emph{front} (Au) side. If $d_{Au}\gg\lambda^S_{Au}$, this signal is neither sensitive to the electromagnetic field of the pump pulse nor to the lattice excitations induced by the pump absorption since their transport through the Au film with the speed of sound takes much longer than the HC transport.

In the following, an experimental approach to separate ballistic and
diffusive contributions to SC directly in the time domain, is demonstrated. A femtosecond experiment in a back-pump-front-probe configuration on a Fe(001)/Au(001) layer stack is employed to realize a time-of-flight-like magneto-optical detection of SC contributions through their different propagation velocities.

Thin films of Fe(001) and Au(001) are grown following \cite{Muehge_APA94,Dekadjevi_PRL01} on optically transparent MgO(001) and examined with the scanning transmission electron microscope (STEM) TITAN 80-300 (FEI, USA) equipped with a spherical aberration corrector and a high angle annular dark field detector. The films grow epitaxially (Fig.~\ref{Fig1}, a, b) with [001]$_{\textrm{Au}}\parallel$ [001]$_{\textrm{Fe}}\parallel$ [001]$_{\textrm{MgO}}$ and [010]$_{\textrm{Au}}\parallel$ [110]$_{\textrm{Fe}}\parallel$ [010]$_{\textrm{MgO}}$ \cite{note1}. The Fe film is magnetized in-plane with the help of an electromagnet.

Second harmonic (SH) generation is used in a back-pump-front-probe scheme for the surface SP detection. The \emph{p}-polarized 35 fs, 800 nm, 40 nJ/pulse, 1.52 MHz output of a cavity dumped Ti:Sapphire oscillator is split at a power ratio of 4:1 into pump and probe pulses. The pump is chopped with a frequency of 500 Hz and the intensity of \emph{p}-polarized SH from the probe is recorded by a two-channel photon counter. The electric field at SH $\vec E(2\omega)=\vec E_{even}\left( {2\omega }\right)+\vec E_{odd}\left( {2\omega } \right)$ consists of even and the odd terms with respect to the reversal of the magnetization $\vec M$ \cite{pan_PRB89}: $\vec E_{even}$ is independent of $\vec M$ and $\vec E_{odd}=\vec \alpha \,M$, where $M$ is the projection of $\vec M$ perpendicular to the optical plane of incidence in the case of \emph{p}-in, \emph{p}-out geometry. Although SC can generate a sizable bulk dipole magneto-induced SH term in semiconductors \cite{Wang_PRL10,Werake_NatPhys10}, in metals only the dipole-allowed surface SH is expected owing to the small $\lambda^S$. Thus, here $\vec M$ is associated with the SP induced by HC at the Au surface.

The surface SP is monitored by the SH intensity
\begin{equation}
    \label{eq:SH intensity}
I\left( {2\omega } \right)\propto \left| {\vec E_{even}\left(
{2\omega } \right)+\vec E_{odd}\left( {2\omega } \right)}
\right|^2\approx k\,M+c
\end{equation}
for $E_{even}\gg E_{odd}$ with $c \equiv E_{even}^2$ and $k \equiv 2\,\vec \alpha \,\vec E_{even}$. Fig.~\ref{Fig1} c, d show hysteresis loops for two selected values of the pump-probe delay $t$. At $t<0$ $I(2\omega)$ is independent of the external magnetic field $H$ applied perpendicular to the plane of incidence, as it is expected for the non-perturbed surface of a paramagnetic metal. However, shortly after the excitation, a dependence $I(H)$ is observed, which closely resembles the hysteresis loop measured at the Fe/MgO interface with the back probe: HC imprint the hysteresis loop of Fe onto the Au surface.

To study the surface dynamics in more details, the SH intensity is measured for a given delay $t$ in the saturating magnetic field of opposite polarities in the presence ($I^{\uparrow \downarrow }\left( {t} \right)$) and in the absence ($I_{0}^{\uparrow \downarrow }$) of excitation. The electronic surface response is characterized by the quantity
\begin{equation}
    \label{eq:D}
\Delta(t)  \equiv \frac{{{E_{even}}\left( t \right) - E_{even}^0}}{{E_{even}^0}} = \sqrt {\frac{{{I^ \uparrow }\left( t \right) + {I^ \downarrow }\left( t \right)}}{{I_0^ \uparrow  + I_0^ \downarrow }}}-1,
\end{equation}
and the transient SP by the SH magnetic contrast
\begin{equation}
    \label{eq:R}
\rho(t)  = \frac{{{I^ \uparrow }\left( t \right) - {I^ \downarrow }\left( t \right)}}{{{I^ \uparrow }\left( t \right) + {I^ \downarrow }\left( t \right)}}=\frac{k(t)M(t)}{c(t)} \propto M\left( t \right).
\end{equation}

\begin{figure} \centering
\includegraphics[width=1\columnwidth]{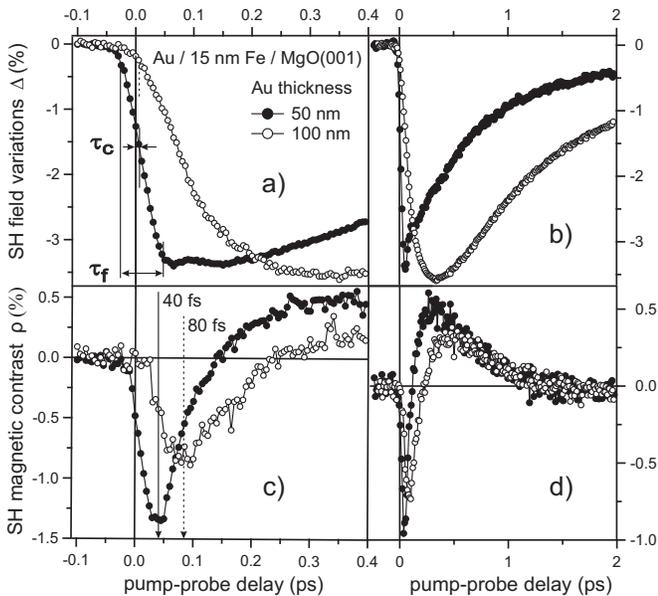}
\caption{\label{Fig2} Pump-induced variation of (a, b) SH electronic contribution $\Delta$ and (c, d) SH magnetic contrast $\rho \propto M$ measured at the Au surface of Au/Fe/MgO(001) structure with indicated $d_{Fe}$ and $d_{Au}$ for the pump fluence of 1~mJ/cm$^2$.}
\end{figure}

Typical results obtained for $\Delta(t)$ and $\rho(t)$ are presented in Fig.~\ref{Fig2}, a-d. Pump-induced variations of the SH field at the surface of 50~nm Au consist in an ultrafast reduction on a time scale of 50~fs followed by a recovery on a 1~ps time scale. The first one can be characterized by the center $\tau_c\approx$10~fs and the width $\tau_f\approx$70~fs of the negative front of $\Delta(t)$ (Fig.~\ref{Fig2}, a). In the case of $d_{Au}$=100~nm the front shifts to $\tau^{\prime}_c\approx$70~fs and broadens to $\tau^{\prime}_f\approx$200~fs; the recovery is also delayed (Fig.~\ref{Fig2}, b). After the excitation, the surface SP builds up and reaches its maximal value at 40~(80)~fs for $d_{Au}$=50~(100)~nm (Fig.~\ref{Fig2}, c). The 40~fs delay agrees with the time required for the traversing of 50~nm of Au by HC propagating with velocities $v^{HC}_{Au}$ close to the Fermi velocity $v^F_{Au}\approx$1.4~nm/fs \cite{Brorson_PRL87}, which is a signature of \emph{ballistic} HC transport. Later on the SH magnetic contrast changes the sign, establishes the second extremum at 0.35 (0.5)~ps and relaxes back to $\rho$=0 at 1~ps time scale for both Au thicknesses (Fig.~\ref{Fig2}, d). To explain this, $v^{HC}_{Au}\ll v^F_{Au}$ should be considered, which can originate from scattering of some fraction of HC leading to \emph{diffusive} transport.

Contributions from \emph{e-e} and \emph{e-phonon} scattering in Au were estimated in \cite{Suarez_PRL95} from experimental results of \cite{Brorson_PRL87,Juhasz_PRB93}: the HC transport is almost purely ballistic for $d_{Au}<100$~nm; for thicker films, a significant diffusive component appears and increases with $d_{Au}$ . Note that the thickness dependence of the ballistic-to-diffusion transport ratio will be affected by the energy spectrum of HC which is different for the HC excitation in Au \cite{Brorson_PRL87,Juhasz_PRB93,Suarez_PRL95} and injection from Fe.

In this picture, $\Delta(t)$ monitors the time profile of HC packet reaching the Au surface with steep "ballistic" front and "diffusive" slope stretched by the carrier scattering. With increasing $d_{Au}$ both the front and the slope become longer (Fig.~\ref{Fig2}, a, b) due to a linear and non-linear increase of the time-of-flight of ballistic and diffusive HC, respectively, and a reduction of ballistic HC percentage.

\begin{figure}
\includegraphics[width=1\columnwidth]{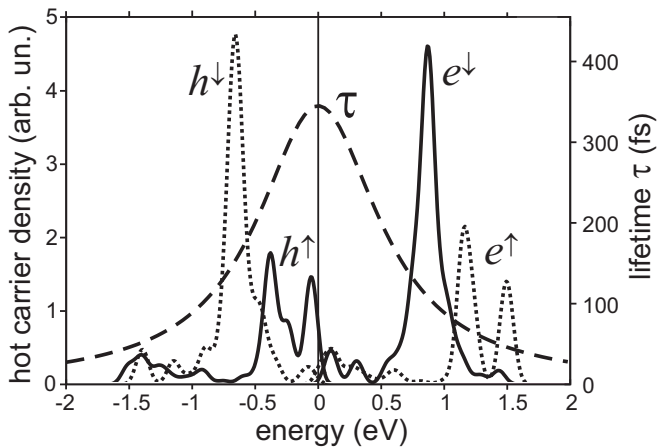}
\caption{\label{Fig3} Calculated density of carriers excited in Fe vs. their energy with respect to the Fermi level: solid (dotted) curves correspond to HC carrying negative (positive) SP. Dashed curve reproduces the HC lifetime from Ref.~\cite{Cao_PRB98}.}
\end{figure}

The magnetic contrast $\rho$ is proportional to the SP induced at the Au surface by two groups of HC excited in Fe: (i) HC$^+$, majority electrons $e^\uparrow$ and minority holes $h^\downarrow$ carrying positive SP and (ii) HC$^-$, $e^\downarrow$ and $h^\uparrow$ with negative SP. The observed non-monotonous behavior of $\rho(t)$ can be explained under the suggestion that the diffusive fraction of HC$^+$ is considerably larger than that of HC$^-$. Then the front of the HC packet will be formed predominantly by HC$^-$ while HC$^+$ will dominate at the slope, which provides the alternating $\rho(t)$ (Fig.~\ref{Fig2}, c). A manifold scattering of HC will lead to their depolarization reducing the SP and leading to $\rho(t)\approx0$ after 1-2~ps (Fig.~\ref{Fig2}, d) while $\Delta(t)\neq0$ (Fig.~\ref{Fig2}, b). This "spin-depolarization" time is estimated to about 1~ps from Fig.~\ref{Fig2}, d.

\begin{SCtable}[1.5]
\begin{tabular}{|c|c|c|c|}
  \hline
   & $v$, nm/fs & $\tau$, fs & $\lambda$, nm \\
  \hline
  $e^\uparrow$ & 0.95 & 60 & 57 \\
  \hline
  $h^\downarrow$ & 0.94 & 150 & 140 \\
  \hline
  $e^\downarrow$ & 1.17 & 120 & 140 \\
  \hline
  $h^\uparrow$ & 0.77 & 300 & 230 \\
  \hline
\end{tabular}

\caption{\label{Table1} Average lifetimes $\tau$, ballistic velocities $v$, and ballistic lengths $\lambda=v \tau$ in Au for different types of spin-polarized hot carriers excited in Fe by 800 nm laser pulse and injected into Au through the Fe/Au interface.}

\end{SCtable}

To verify this explanation, the energy and momentum distribution of HC excited in Fe was calculated based on density functional theory as implemented in the WIEN2k package \cite{Wien2k} using the linearized augmented plane wave method and a generalized gradient approximation \cite{Perdew_PRL96} to the exchange and correlation potential. Fig.~\ref{Fig3} shows the spectrum of excited HC after the momentum averaging. Upon the injection of HC across the Fe(001)/Au(001) interface the energy and the momentum component parallel to the interface are conserved \cite{note1}. This allows to calculate the normal-to-interface momentum component of the injected HC and the corresponding $v^{HC}_{Au}$. The velocities were calculated based on density functional theory using band structures of bulk Au obtained within the Vienna Ab-Initio simulation package (VASP) \cite{Kresse_JPCM94} with the projector augmented waves basis sets \cite{Bloechl_PRB94,Kresse_PRB99}.

The results presented in Table~\ref{Table1} show that the difference between the \emph{ballistic} velocities $v$ is not significant. On the other hand, the scattering rate of HC in Au is characterized by the \emph{lifetime} $\tau^{HC}_{Au}$ or by the \emph{ballistic propagation length} $\lambda^{HC}_{Au}$. The energy-dependent $\tau^{HC}_{Au}$ \cite{Cao_PRB98} is depicted in Fig.~\ref{Fig3}. This approximation is based on Fermi liquid theory but, in addition to the \emph{e-e} scattering, takes into account the \emph{e-phonon} scattering phenomenologically as well as other contributions. The curve was obtained as a fit to the results of time-resolved two photon photoemission (2PPE) experiments \cite{Cao_PRB98} and thus can underestimate $\tau^{HC}_{Au}$ due to the transport of HC out of the volume probed in 2PPE. Also in the vicinity of the Fermi level the vanishing \emph{e-phonon} scattering will increase the lifetime. Although 2PPE is not sensitive to elastic scattering dominating the transport experiments \cite{jansen_JPhysD03}, $\tau^{HC}_{Au}$ from Fig.~\ref{Fig3} can be used here due to very low defect concentration in epitaxial Fe/Au structures vs. polycrystalline metal films from Ref.~\cite{jansen_JPhysD03} (see Fig.~\ref{Fig1}, b) \cite{note1}. Average lifetimes and corresponding ballistic lengths collected in Table~\ref{Table1} show that HC excited in the minority sub-band of Fe give a moderate diffusive contribution to both HC$^+$ and HC$^-$ transport from $h^\downarrow$ and $e^\downarrow$, respectively, for the studied $d_{Au}$. In contrast, HC generated in the majority sub-band of Fe give an almost pure ballistic contribution to the HC$^-$ and an essentially diffusive one to the HC$^+$ transport supporting the scenario sketched above (see Fig.~\ref{Fig3} for illustration). Thus, the results presented in Fig.~\ref{Fig2} can be indeed attributed to the competition between ballistic and diffusive transport of spin-polarized HC through the Au layer. This qualitative picture of diffusive vs. ballistic HC transport is likely to be oversimplified: strictly speaking, a packet of HC propagating in the superdiffusive regime \cite{battiato_PRL10} should be considered. With that, however, the faster HC ("more ballistic" part) will have, on average, a negative and the slower HC ("more diffusive" contribution) a positive SP.

\begin{figure} \centering
\includegraphics[width=1\columnwidth]{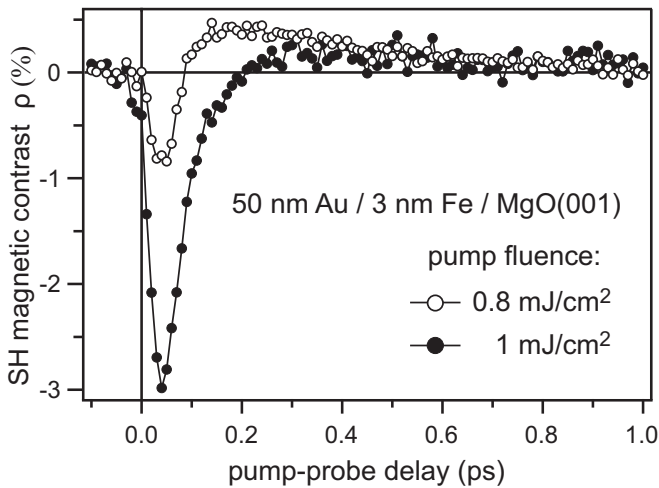}
\caption{\label{Fig4} Pump-induced variation of SH magnetic contrast $\rho \propto M$ measured at the Au surface of Au/Fe/MgO(001) structure with $d_{Fe}=$3~nm, $d_{Au}=$50~nm for indicated pump fluences.}
\end{figure}

For applications, it could be rewarding to have a pure HC$^-$ or HC$^+$ contribution since it would represent a single-polarity SC pulse. In the general case, interface transmission coefficients $\sigma_{Fe/Au}^{HC^+}\neq\sigma_{Fe/Au}^{HC^-}$ since HC$^+$ and HC$^-$ have different energies. They depend on the interface properties and this dependence might be different for $\sigma_{Fe/Au}^{HC^+}$ and $\sigma_{Fe/Au}^{HC^-}$. Regarding to that, exemplary $\rho(t)$ curves measured at a sample with smaller Fe thickness are shown in Fig.~\ref{Fig4}. Employing a 0.8~mJ/cm$^2$ pump fluence, the result is similar to that presented in Fig.~\ref{Fig2}. Upon an increase of the pump fluence up to 1~mJ/cm$^2$ (i) the amplitude of the negative peak increases more than in three times and (ii) the positive feature becomes negligibly small. Since both effects are irreversible with respect to the pump fluence, they are attributed to a laser-induced modification of the Fe/Au interface leading to strong increase of $\sigma_{Fe/Au}^{HC^-}$ and generation of 100~fs single-polarity SC pulses. Estimations based on the absorbed pump fluence show that the generated SC corresponds to the spin polarized carrier transport on the order of $10^8-10^9$ A/cm$^{2}$ or $10^{14}-10^{15}$ HC/cm$^2$/pulse or 0.1-1 HC/atom in the surface layer.

In future experiments, a third layer can be produced on top of the gold film, which is a FM metal, FM nanostructure, FM-ordered molecular layer, etc. A magneto-optical probe will monitor the magnetization dynamics excited in this layer by a short SC pulse. Since the Au film blocks the pump pulse and delays the acoustic transport, the excitation by spin-polarized HC can be monitored selectively. This might help to unravel the origin of ultrafast magnetization dynamics.

In conclusion, a time-of-flight-like investigation of spin transport is demonstrated in Au/Fe/MgO(001). Spin current pulses formed by spin-polarized hot carriers excited in Fe, are detected at the Au surface by optical second harmonic generation. The dependence of velocity of carriers propagating in the superdiffusive regime, on their energy/spin-polarization distribution, lead to an alternating transient surface spin polarization.

Financial support by the DFG through ME 3570/1-1 and SFB 668 (TP A3) and partly by the Russian Foundation for Basic Research, is acknowledged.



\begin{thebibliography}{10}

\bibitem{Hillebrands2006}
 in {\em Spin Dynamics in Confined Magnetic Structures III}, Vol.~101 of {\em
  Topics Appl. Phys.}, edited by B. Hillebrands and A. Thiaville
  (Springer-Verlag, Berlin, 2006).

\bibitem{kirilyuk_RevModPhys10}
A. Kirilyuk, A.~V. Kimel, and T. Rasing, Rev. Mod. Phys. {\bf 82},  2731
  (2010).

\bibitem{koopmans_PRL05}
B. Koopmans, J.~J.~M. Ruigrok, {F. Dalla Longa}, and W.~J.~M. de~Jonge, Phys.
  Rev. Lett. {\bf 95},  267207  (2005).

\bibitem{koopmans_NatMat10}
B. Koopmans {\it et~al.}, Nature Materials {\bf 9},  259  (2010).

\bibitem{bigot_NatPhys09}
J.-Y. Bigot, M. Vomir, and E. Beaurepaire, Nature Physics {\bf 5},  515
  (2009).

\bibitem{zhang_NatPhys09}
G.~P. Zhang {\it et~al.}, Nature Physics {\bf 5},  499  (2009).

\bibitem{battiato_PRL10}
M. Battiato, K. Carva, and P.~M. Oppeneer, Phys. Rev. Lett. {\bf 105},  027203
  (2010).

\bibitem{jansen_JPhysD03}
R. Jansen, J. Phys. D: Appl. Phys. {\bf 36},  R289  (2003).

\bibitem{Malinowski_NatPhys08}
G. Malinowski {\it et~al.}, Nature Physics {\bf 4},  855  (2008).

\bibitem{Brorson_PRL87}
S.~D. Brorson, J.~G. Fujimoto, and E.~P. Ippen, Phys. Rev. Lett. {\bf 59},
  1962  (1987).

\bibitem{Juhasz_PRB93}
T. Juhasz {\it et~al.}, Phys. Rev. B {\bf 48},  15488  (1993).

\bibitem{Ralph_JMMM08}
D.~C. Ralph and M.~D. Stiles, J. Magn. Magn. Mat. {\bf 320},  1190  (2008).

\bibitem{Muehge_APA94}
T. Mühge {\it et~al.}, Appl. Phys. A {\bf 59},  659  (1994).

\bibitem{Dekadjevi_PRL01}
D.~T. Dekadjevi {\it et~al.}, Phys. Rev. Lett. {\bf 86},  5787  (2001).

\bibitem{note1}
The epitaxial growing is important to obtain a well-defined structure suitable
  for \emph{ab-initio} analysis and minimize the diffusive contribution to the
  HC transport originating from the scattering on lattice inhomogeneities
  \cite{Juhasz_PRB93}.

\bibitem{pan_PRB89}
R.-P. Pan, H.~D. Wei, and Y.~R. Shen, Phys. Rev. B {\bf 39},  1229  (1989).

\bibitem{Wang_PRL10}
J. Wang, B.-F. Zhu, and R.-B. Liu, Phys. Rev. Lett. {\bf 104},  256601  (2010).

\bibitem{Werake_NatPhys10}
L.~K. Werake and H. Zhao, Nature Physics  (2010).

\bibitem{Suarez_PRL95}
C. Suárez, W.~E. Bron, and T. Juhasz, Phys. Rev. Lett. {\bf 75},  4536  (1995).

\bibitem{Cao_PRB98}
J. Cao {\it et~al.}, Phys. Rev. B {\bf 58},  10948  (1998).

\bibitem{Wien2k}
P. Blaha {\it et~al.}, {\em Wien2k, An Augmented Plane Wave + Local Orbitals
  Program for Calculating Crystal Properties} (Vienna University of Technology,
  Austria, 2001).

\bibitem{Perdew_PRL96}
J.~P. Perdew, K. Burke, and M. Ernzerhof, Phys. Rev. Lett. {\bf 77},  3865
  (1996).

\bibitem{Kresse_JPCM94}
G. Kresse and J. Hafner, J. Phys.: Condens. Matter {\bf 6},  8245  (1994).

\bibitem{Bloechl_PRB94}
P.~E. Blöchl, Phys. Rev. B {\bf 50},  17953  (1994).

\bibitem{Kresse_PRB99}
G. Kresse and D. Joubert, Phys. Rev. B {\bf 59},  1758  (1999).

\end{thebibliography}

\end{document}